\documentclass[unnumsec,webpdf,contemporary,large]{oup-authoring-template}%

\graphicspath{{Fig/}}

\theoremstyle{thmstyleone}%
\theoremstyle{thmstyletwo}%
\theoremstyle{thmstylethree}%

\usepackage{natbib}
\usepackage{float}

\begin{document}

\journaltitle{Journal Title Here}
\DOI{DOI HERE}
\copyrightyear{2022}
\pubyear{2019}
\access{Advance Access Publication Date: Day Month Year}
\appnotes{Paper}

\firstpage{1}


\title[DNABERT-Cap]{Predicting Transcription Factor Binding Sites using Transformer based Capsule Network}

\author[1,2,$\ast$]{Nimisha Ghosh}
\author[2]{Daniele Santoni}
\author[3]{Indrajit Saha}
\author[2]{Giovanni Felici}

\authormark{Nimisha Ghosh et al.}

\address[1]{\orgdiv{Department of Computer Science and Information Technology}, \orgname{Institute of Technical Education and Research, , Siksha `O' Anusandhan (Deemed to be University)}, \orgaddress{\street{Bhubaneswar}, \postcode{751030}, \state{Odisha}, \country{India}}}
\address[2]{\orgdiv{Institute for System Analysis and Computer Science “Antonio Ruberti”}, \orgname{National Research Council of Italy}, \orgaddress{\street{Via dei Taurini}, \postcode{00185}, \state{Rome}, \country{Italy}}}
\address[3]{\orgdiv{Department of Computer Science and Engineering}, \orgname{National Institute of Technical Teachers' Training and Research}, \orgaddress{\street{Kolkata}, \postcode{700106}, \state{West Bengal}, \country{India}}}

\corresp[$\ast$]{Corresponding author. \href{ghosh.nimisha@gmail.com}{ghosh.nimisha@gmail.com}}

\received{Date}{0}{Year}
\revised{Date}{0}{Year}
\accepted{Date}{0}{Year}


\abstract{
\textbf{Motivation}: Prediction of binding sites for transcription factors is important to understand how they regulate gene expression and how this regulation can be modulated for therapeutic purposes. Although in the past few years there are significant works addressing this issue, there is still space for improvement. In this regard, a transformer based capsule network viz. DNABERT-Cap is proposed in this work to predict transcription factor binding sites mining ChIP-seq datasets. DNABERT-Cap is a bidirectional encoder pre-trained with large number of genomic DNA sequences, empowered with a capsule layer responsible for the final prediction. \\
\textbf{Results}: The proposed model builds a predictor for transcription factor binding sites using the joint optimisation of features encompassing both bidirectional encoder and capsule layer, along with convolutional and bidirectional long-short term memory layers. To evaluate the efficiency of the proposed approach, we use a benchmark ChIP-seq datasets of five cell lines viz. A549, GM12878, Hep-G2, H1-hESC and Hela, available in the ENCODE repository. The results show that the average area under the receiver operating characteristic curve score exceeds 0.91 for all such five cell lines. DNABERT-Cap is also compared with existing state-of-the-art deep learning based predictors viz. DeepARC, DeepTF, CNN-Zeng and DeepBind, and is seen to outperform them.\\
\textbf{Availability and implementation}: The datasets used in this work along with code are available at https://github.com/NimishaGhosh/DNABERT-Cap/tree/main.\\
\textbf{Contact}: ghosh.nimisha@gmail.com}


\maketitle

\section{Introduction}
Transcription Factors (TFs) are proteins that bind to certain genomic sequences and influence a wide range of cellular functions \citep{LATCHMAN1997,Karin1990}. TFs bind to DNA-regulatory sequences which are known as Transcription Factor Binding Sites (TFBSs), typically of size 4-30 bp \citep{Tompa2005, Tan2016, Qu2019} and modulate the gene transcription along with playing important role in cellular processes \citep{Alexandrov2010, Li2001, WILKINSON2017}. The correct prediction of TFBSs is indeed crucial to characterise certain functional aspects of the genome as well as explain the organisation of specific sequence expression in complex organisms \citep{LAMBERT2018, BASITH2018, Shen2019}. High-throughput sequencing technology has led to the generation of a huge amount of experimental data about TFBS, like JASPAR \citep{Fornes2019}, TRANSFAC \citep{Matys2006} etc. However, identifying TFBSs using experimental methods is very slow and expensive, thereby leading to the development of computational methods to identify TFBSs encompassing large amount of data. 

Initially, many researchers proposed machine learning methods to identify TFBSs. In this regard, Wong et al. \citep{Wong2013} put forth kmerHMM to identify TFBS. In this method, Hidden Markov Model (HMM) is trained for the underlying motif representation and subsequently belief propagation is used to extract multiple motifs from HMM. To predict DNA binding site, Ghandi et al. \citep{Ghandi2014} proposes gkm-SVM which uses a tree for the calculation of the kernel matrix. However, traditional machine learning models usually rely on manual feature extraction and they have problem processing large-scale datasets. In recent times, there have been a number of  deep learning models specifically developed for computer vision \citep{He2016,he2016deep} and natural language processing \citep{devlin2019}. Similar models have also been applied to solve problems in computational biology and bioinformatics  \citep{Zhao2021,Min2021,Liu2021}. Deep neural network based methods such as DeepBind~\citep{Alipanahi2015} and DeepSEA~\citep{zeng2016} shows competitively better results as compared to traditional methods like Markov model, support vector machines,  hierarchical mixture models, discriminative maximum conditional likelihood and random forests. DeepBind~\citep{Alipanahi2015} demonstrates the capabilities of deep learning to assess sequence specificity from experimental data. It offers a scalable, adaptable and integrated calculating technique for finding patterns. DeepBind is also the first technology to ever address the demand for precise modelling of protein target binding motifs. A long-short term recurrent convolutional network called DeeperBind~\citep{hassanzadeh2016} is used to anticipate the specificities of how proteins will bind to DNA probes. In order to effectively account for the contributions provided by various sub-regions in DNA sequences, DeeperBind can describe the positional dynamics of probe sequences. It can also be trained and evaluated on datasets with sequences of different lengths. Quang et al.~\citep{Quang2016} propose DanQ that combines CNNs and bidirectional long short-term memory network (BiLSTM) to predict binding sites. Zeng et al.~\citep{zeng2016} uses multiple CNN architectures for the prediction of DNA sequence binding using an extensive collection of transcription factor datasets. In order to split the DNA binding sequence into overlapping pieces and predict TFBS, Farrel et al.~\citep{Farrel2017} present an effective pentamer approach. In~\citep{Qin2017}, Qin et al. propose TFImpute to predict cell-specific TFBS on ChIP-seq data. This method incorporates TFs and cell lines into continuous vectors that are used as inputs to the model. DeepSNR as proposed by Salekin et al.~\citep{Salekin2018} uses CNN-Deconvolutional model to predict transcription factor binding location at single nucleotide resolution. DeepFinder~\citep{Lee2018} is an enhanced three-stage DNA motif prediction for the large-scale pattern analysis that uses TFBS-associated deep learning neural networks to build the motif model. For data on imbalanced DNA-protein binding sites, Zhang et al.~\citep{ZHANG2019} suggest a new prediction approach. This technique employs Bootstrap algorithm to undersample the negative data, while adaptive synthesis is used to oversample positive data. To further capture long-term relationships between DNA sequence motifs, DeepSite~\citep{Zhang2020} uses CNN and BiLSTM. Apart from considering sequence dependencies, filtering out valid information in huge data and precisely locating motif information in the imbalanced data are also essential questions addressed in~\citep{Zhang2020}. Yang et al.~\citep{Yang2019} use deep neural networks along with binomial distribution to enhance motif prediction in the human genome to help with TFBS identification and motif prediction accuracy. In~\citep{Chen2021}, Chen et al. use deep learning to develop TF binding prediction tool known as DeepGRN. The first part of the model is a convolutional layer while the BiLSTM nodes are recurrent units. Multi-scale convolution along with LSTM (MCNN-LSTM) are the choice made in~\citep{Bao2019} to accurately predict TFBSDeepTF. Their results show that MCNN-LSTM outperforms several existing TFBS predictors. Zhang et al.~\citep{ZHANG2021} combine convolutional autoencoder with convolutional neural network (CAE-CNN) to predict TFBS and use a gated unit to understand the features better. 
Their primary contribution lies in the integration of supervised and unsupervised learning methods to predict TFBS. In~\citep{Jing2022}, Jing et al. use meta learning based CNN method (MLCNN) to predict TFBS. The performance of MLCNN shows that it is competitively better than other state-of-the-art CNN methods. A hybrid convolutional recurrent neural network (CNN/RNN) architecture known as CRPTS is proposed in~\citep{WANG2021} to predict TFBSs by combining DNA sequence and DNA shape features. Cao et al.~\citep{Cao2022} propose DeepARC which combines convolutional neural network (CNN)  and recurrent neural network (RNN) to predict TFBS. DeepARC uses an encoding method by combining OneHot encoding and DNA2Vec. This method shows promising results in terms of AUC for benchmark datasets. However, DeepARC lacks in adopting efficient encoding policies.

Thus, it can be very well concluded that deep learning is widely applied in the prediction of TFBSs. Although, many aspects of deep learning are well explored in the context of TFBS prediction, works based on transformers are quite limited. Motivated by the literature, in this work we develop DNABERT-Cap built on top of pre-trained DNABERT model to predict TFBSs. Initially, we consider the positive set which are cell line specific DNA sequences with TFBSs while the negative set is positive shuffled sequences conserving dinucleotide frequency. These DNA sequences are divided into \textit{k}-mers which are then sequentially passed through DNABERT, a Convolutional Layer (CL),  a bidirectional Long-Short Term Memory (BiLSTM) layer and finally to a capsule layer,  in order to make effective prediction of TFBSs. 

This transformer based capsule network viz. DNABERT-Cap uses DNABERT to provide the embedding of DNA sequences and apply CL with BiLSTM along with capsule layer to create an effective prediction model.
DNABERT is a bidirectional encoder pre-trained with a large number of DNA sequences. It uses positional embedding to understand the semantic relationship among the \textit{k}-mers of a DNA sequence. Along with DNABERT, the convolutional layer provides a further abstract representation of the \textit{k}-mers taking spatial correlations into account while the BiLSTM layer captures the context of \textit{k}-mers to learn long term dependencies in a DNA sequence. Finally, the addition of a capsule layer improves the efficiency of a predictor by encoding spatial patterns to process the relationship between a part and the whole of the sequence through capsules and dynamic routing.
Extensive experimental results show that the proposed method performs better than existing state-of-the-art methodologies. It is worth mentioning over here that capsule network is already used in natural language processing (NLP) for text \citep{KIM2020, CHEN2020}, tweet act classification \citep{Saha2020} as well as in bioinformatics \citep{Zhang2021_new, Cheng2021} with competitive results.

\section{Materials and Methods}
In this section, data preparation is elaborated, followed by the discussion of the pipeline of the proposed work. 
\subsection{Data Preparation}
In this work, the benchmark dataset from Encyclopedia of DNA Elements (ENCODE) \citep{encode2012} is used to acquire the TFBS data analysed by ChIPseq method. This data is used to train and test the proposed model. Data preprocessing is the same as considered in \citep{zeng2016}, where the positive samples with 101 bps are generated in the centre of each ChIP-seq peak, while the negative samples are obtained by recombining the positive sequence conserving dinucleotide sequences. The positive and negative samples are then distinguished based on the presence or absence of TFBSs in a sequence.
For experimental purposes and based on availability of resources, 500,000 sequences for each of the five cell lines viz. A549, GM12878, Hep-G2, H1-hESC and Hela are selected randomly from the ChIP-seq datasets. This random selection takes care of a fair balance between positive and negative sequences. Each dataset is divided into 70\% training, 20\% test and 10\% validation sets. All the experiments are conducted on machines with NVIDIA GA100 GPUs. 

\subsection{DNABERT and Capsule Network}
Before delving into the pipeline of the work, a brief discussion on DNABERT model and capsule network is put forth.
\subsubsection{DNABERT}
DNABERT \citep{Ji2021} is a pre-trained bidirectional encoder representation which captures intricacies of DNA sequences based on up and downstream nucleotide contexts. A set of sequences divided into \textit{k}-mer tokens of appropriate sizes are provided as input to DNABERT. Each sequence is represented as a matrix $\mathcal{X}$ where the tokens are embedded into a numerical vectors. Such matrix captures contextual information of a sequence by executing multi-head self-attention mechanism on $\mathcal{X}$:
\begin{equation}
    multihead\mathcal{X} = Concatenation(head_1,\dots,head_h)\mathcal{W}^O
    \label{bert1}
\end{equation}
where 
\begin{equation}
    head_i = softmax\left(\frac{\mathcal{X}\mathcal{W}^Q_i\mathcal{X}\mathcal{W}^K_iT}{\sqrt{d_k}}\right).\mathcal{X}\mathcal{W}^V_i
    \label{bert2}
\end{equation}
Here, all $\mathcal{W}$s are parameters learned during training of the model. Equations (\ref{bert1}) and (\ref{bert2}) are performed $T$ times, where $T$ is the number of layers. 

\subsubsection{Capsule Networks}
In order to derive local patterns from a vector sequence, CNN builds convolutional feature detectors \citep{Saha2020}. The most noticeable patterns are then chosen using max-pooling. However, CNN may lose many important information during such pooling process and perform poorly on problems characterised by positional invariance. In contrast, approaches that do not take spatial relationships into account perform flawlessly when making inferences for local patterns;  however, they cannot encode rich structures that may be present in a sequence. In this regard, capsule networks \citep{hinton2018} help to improve the efficiency to encode spatial patterns to include knowledge about the part-whole relationships. Each capsule is a group of neurons where the input and output are both vectors. These group of neurons work together to recognise specific features or patterns. They have an iterative dynamic routing algorithm \citep{Sabour2017} which helps to decide the most important features from lower to higher layers. As a result, capsule networks generalise a particular class instead of memorising every viewpoint variant of the class, thereby becoming invariant to the viewpoint and showing improved performance as compared to CNN. 
\subsection{Pipeline of the Work}
\begin{figure*}
	\centerline{
		\includegraphics[height=2.0in,width=6.0in]{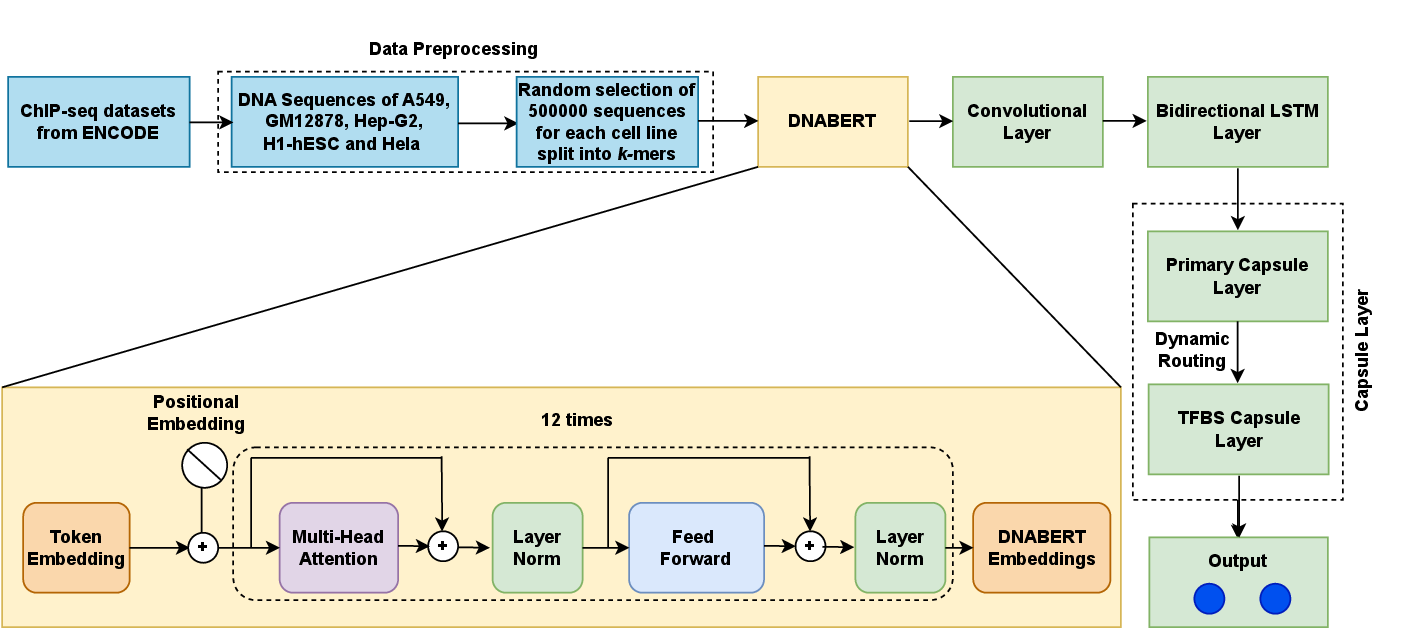}}
	\caption{Depiction of the pipeline of the methodology of DNABERT-Cap. The first row represents the main sequences of the components, while the yellow box below is an explosion of the DNABERT sequence. The last column of boxes, in green, is where the capsule layer is out in force (two capsules are represented with blue dots in the output layer).
 }
	\label{pipeline}
\end{figure*}

The pipeline of the work is depicted in Fig. \ref{pipeline}. Initially, the DNA sequences for each cell line are fed to the DNABERT model in the form of \textit{k}-mer tokens (\textit{k}=6 in this case as \citep{Ji2021} reports that \textit{6}-mers show the best performance). For each input sequence of tokens, DNABERT returns an embedding. Let $W_A \in \mathbb{R}^{d\times l}$ be the weight matrix for each such embedding, where $l$ is the length of a sequence and $d$ is the dimension of each token representation. Each weight matrix is then passed through a series of layers to obtain the best possible sequence representation for the classification of such sequences as Transcription Factor Binding Site. The subsequent layers are as follows:
\begin{enumerate}
    \item \textbf{Convolutional Layer:} The output from DNABERT is provided as an input to the convolutional layer followed by a pooling layer to extract the feature map of a sequence. This is represented as:
    
    \begin{equation}
    \alpha_i = W_b*d_i+b
    \label{eq:1}
\end{equation}
\begin{equation}
    \hat{\alpha}_i = pool(\alpha_i)
    \label{eq:2}
\end{equation}
Here, output feature map $\alpha_i$ is produced by kernel $d_i$ with bias $b$ by applying convolution and pooling $pool(.)$. $\eta$ such feature maps are then combined to form a $\eta$-channel layer as
\begin{equation}
    A = [\hat{\alpha}_1,\hat{\alpha}_2,\dots,\hat{\alpha}_\eta]
    \label{eq:3}
\end{equation}
This layer thus helps in understanding the importance of a \textit{k}-mer token in a sequence by concentrating on the feature map.

\item \textbf{Bidirectional LSTM Layer:} In order to learn semantic dependency, feature vector as obtained from the previous layer is passed through a bidirectional Long Short Term Memory (BiLSTM). The long term dependencies in a sequence is captured using such a BiLSTM by sequentially encoding the feature maps into hidden states \citep{Saha2020}.
In this regard, $\eta$-channel feature vector $A$ is passed through a BiLSTM.
\begin{equation}
    \overrightarrow{h}_t = \overrightarrow{LSTM}(\hat{\alpha}_\eta,h_{t-1})
    \label{eq:4}
\end{equation}
\begin{equation}
    \overleftarrow{h}_t = \overleftarrow{LSTM}(\hat{\alpha}_\eta,h_{t+1})
    \label{eq:5}
\end{equation}
Each feature map is mapped to forward and backward hidden states. This helps in retaining the context-sensitive nature of the tokens. The final hidden state matrix is defined as: 
\begin{equation}
    HS = [h_1,h_2,\dots,h_t]
    \label{eq:6}
\end{equation}
where $HS\in\mathbb{R}^{t\times2dim}$, $dim$ is the number of hidden state.
Thus, BiLSTM is important for capturing the context of \textit{k}-mers in a DNA sequence.

\item \textbf{Primary Capsule Layer:} The primary capsule layer was originally introduced to handle the drawbacks of conventional CNN by replacing the scalar outputs with vector-output capsules to preserve the local order and semantic representations of tokens. Keeping this in context, the features as obtained from the previous layers in the form of vectors are fed into the primary capsule layer. By sliding over the hidden states $HS$ (generated in the previous layer), each kernel $d_i$ generates a sequence of capsules $cap_i$ of dimension $dim$, thereby creating a channel $C_i$.
\begin{equation}
    C_i = S(d_i*HS+b)
    \label{eq:7}
\end{equation}
Here, $S$ is the squash function and $b$ represents the capsule bias weight parameter. 
This layer thus captures the local ordering of \textit{k}-mers in a sequence and its corresponding semantic representations.

\item \textbf{Dynamic Routing Between Capsules:} The primary idea behind dynamic routing \citep{Sabour2017} is building non-linear map iteratively. This ensures that a suitable capsule in the next layer is strongly connected to a lower level capsule. Moreover, pooling function of the traditional convolution layer which normally removes the location information is replaced with dynamic routing technique leading to a robust network. To ensure that the length of a capsule is within {[0,1]}, a nonlinear squashing function is applied as given in Equation (\ref{eq:9})

\begin{equation}
   v_y = \frac{||s_y||^2 }{1+||s_y||}\frac{s_y}{|||s_y||^2}
   \label{eq:9}
\end{equation}
Here, $v_y$ is the output vector of capsule $y$ while $s_y$ is its total input. $s_y$ is a weighted sum over all prediction vectors $\hat{u}_{y|x}$ calculated by capsule $x$ and transferred to capsule $y$. $\hat{u}_{y|x}$ is calculated by multiplying previous layer capsule output $u_x$ by $W_{xy}$ (weight matrix). This process helps a capsule network to capture the relation between a subpart and the entire sequence.

\begin{equation}
    \hat{u}_{x|y} = W_{xj}u_x
    \label{eq:10}
\end{equation}
\begin{equation}
    s_y = \sum_x c_{xy}\hat{u}_{y|x}
    \label{eq:11}
\end{equation}
Here, $c_{xy}$ are the coupling coefficients calculated by the dynamic routing algorithm. $c_{xy}$ is computed as a softmax $b_{xy}$ which is a log prior probabilities between capsules $x$ and $y$.
\begin{equation}
    c_{xy} = \frac{exp(b_{xy})}{\sum_k exp(b_{xk})}
    \label{eq:12}
\end{equation}

The initial coupling coefficients are refined iteratively, based on $b_{xy}$,  measuring the agreement between $v_y$ and $\hat{u}_{y|x}$. Such agreement can be calculated as: $a_{xy} = v_y.\hat{u}_{y|x}$, where $a_{xy}$ is a scalar product. $b_{xy}$ is updated as:
\begin{equation}
    b_{xy} = b_{xy} + \hat{u}_{y|x}.v_y
    \label{eq:12}
\end{equation}

This entire procedure reflects the dynamic routing for all capsule $s$ in layer $P$ and capsule $y$ in layer $P+1$. In this work, dynamic routing algorithm helps to decide the importance and agreement of tokens for a specific task by learning the importance of \textit{k}-mer tokens in a sequence.

\item \textbf{TFBS Capsule Layer:} In this layer, the TFBS capsules are responsible for detecting the TFBS of a given DNA sequence. The sequence vector of the primary capsule is carried forward to the TFBS capsule layer which generates one vector for each of the TFBS class capsule; one for the class which exhibits the presence of TFBS; and another one depicting the absence of the same.

\item \textbf{Output:} The output from TFBS capsule layer has two class capsules (denoted by two blue circles in Fig. \ref{pipeline}) generating outputs with two vectors encoding various properties of features and the lengths are the probabilities that the corresponding class is present in the input data.
In order to improve the separation between the two class capsules, separate margin loss \citep{Sabour2017} $L_b$ is used in this work.

\begin{equation}\small
    L_b = G_b max(0,m^+ - ||v_k||)^2 + \lambda (1-G_b) max (0,m^- - ||v_b||)^2
\end{equation}
where $v_b$ is the capsule for class $b$. $G_b = 1$ iff class $b$ is the ground truth and $m^+$ = 0.9 and $m^-$ = 0.1. $\lambda$ is used to tune the weight of an absent class.
\end{enumerate}

\subsection{Hyperparameters}
The pre-trained DNABERT \citep{Ji2021} model has 12 transformer layers and 768 hidden units along with 12 attention heads in each layer. For both  convolutional and BiLSTM layers, the number of units is 64 while the kernel size used is 2 for convolutional layer. Furthermore, a dropout value of $0.3$ and a capsule length of 16 are considered. The dynamic routing algorithm with 3 iterations provided the optimum results and Adam optimiser \citep{kingma2015} is used for all the experiments. All these parameters are considered after conducting thorough experiments.

\section{Results}
DNABERT-Cap is devised for predicting TFBSs in DNA sequences. The reported results are an average of multiple runs on 500,000 randomly chosen sequences for each cell line viz. A549, GM12878, Hep-G2, H1-hESC and Hela. The reported results are conducted on several such randomly generated datasets and they confirm the efficacy of DNABERT-Cap. We also have conducted similar experiments on such other datasets which show similar results (available on request). To show the effectiveness of the proposed model, it is compared with some baselines as well as other state-of-the-art approaches. The performance measures for such comparisons are mentioned next.

\subsection{Performance Metrics}
The different performance parameters as used in this work are Accuracy, Recall, Specificity Mathew's Correlation Coefficient (MCC) and area under the receiver operating characteristic curve (AUC) \citep{Cao2022}.

\subsection{Performance Comparison with Baselines}
To show the efficacy of the proposed model, it is compared with the following baseline models:
\begin{enumerate}
    \item Fine-tuned DNABERT model (Baseline-1): For comparison purposes, the original DNABERT model is fine-tuned with our dataset where we have added a classification layer on top of DNABERT.
    \item DNABERT+CL+BiLSTM+CE (Baseline-2): The usage of capsule layer to calculate the loss is removed from this baseline and instead categorical cross entropy is used as the loss function.
    \item DNABERT+CL+Capsule Layer (Baseline-3): In this baseline, the BiLSTM layer is removed to provide the comparison.
    
\end{enumerate}
The results in terms of accuracy, recall, specificity, MCC and AUC are reported in Table \ref{tab1}. As can be observed from Table \ref{tab1}, DNABERT-Cap provides competitive results if compared to all the other baselines. When compared to the fine-tuned DNABERT model, DNABERT-Cap shows an improvement of around 5\% in terms of accuracy for cell line A549. Similar improvements can be observed for other metrics as well for other cell lines. Baseline 2 which includes DNABERT, convolutional and BiLSTM layer without the advantage of capsule layer also shows quite competitive results as compared to baseline-1. This is true for all the considered cell lines considered in this work.
As compared to baseline-2, where the advantage of capsule layer is not taken into account - the proposed model shows improvement in terms of accuracy, specificity, MCC and AUC of around 4\% for A549. However, in terms of recall, baseline-3 shows better performance. In this regard, to determine which baseline has the best performance, AUC is a better parameter \citep{GHOSH2021}. As can be seen from the results, DNABERT-Cap has an improved performance in terms of AUC. This improved performance is reflected for all the other cell lines as well. It is worth noting here that as compared to baseline-3, the overall improved performance of the proposed model can be attributed to the addition of a BiLSTM layer. All the experiments are performed considering a confidence level of 95\%.
Supplementary Figure S1 shows the area under the curve for all the five cell lines for one of the multiple runs of the proposed DNABERT-Cap where the AUC values of A549, GM12878, Hep-G2, H1-hESC and Hela are 0.925, 0.913, 0.930, 0.914 and 0.917 respectively. In the figures, no skill is represented at (0.5,0.5). Please note that for the ease of understanding and explanation, out of multiple runs, curves with similar AUC values to those as reported in Table \ref{tab1} are provided over here.

\subsection{Performance Comparison with State-of-the-Art Predictors}
\begin{figure*}
	\centerline{
		\includegraphics[height=2.0in,width=4.0in]{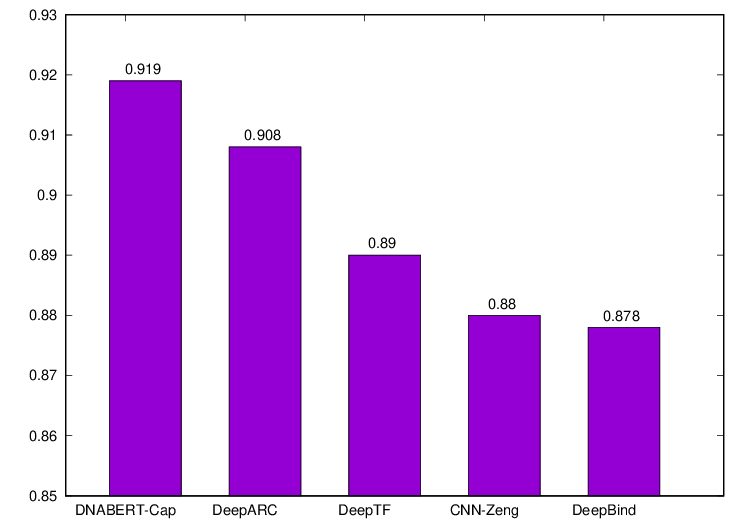}}
	\caption{Chart depicting the \textit{area under the ROC curve} of DNABERT-Cap and the four selected state-of-the-art predictors. DNABERT-Cap has the best AUC value.}
	\label{AUC}
\end{figure*}

In order to analyse the performance of the proposed DNABERT-Cap, we compare it with several other prediction methods viz. DeepARC \citep{Cao2022}, DeepTF \citep{Bao2019}, CNN-Zeng \citep{HZeng2016} and DeepBind \citep{Alipanahi2015} considering the same cell lines as mentioned previously. Table \ref{tab2} reports the results of these comparisons considering the average of the five cell lines. As evident from the table, the proposed model has the best predictive performance among all the other state-of-the-art approaches in terms of accuracy, specificity, MCC and AUC. With respect to recall, DeepARC shows the best performance. As compared to DeepARC, DNABERT-Cap has an improved performance of 1.11\%, 0.01\%, 2.2\% and 1.10\% for accuracy, specificity, MCC and AUC respectively. Fig. \ref{AUC} reports the AUC of each method for the five cell lines. It is to be noted that \citep{Zhang2021_new, Cheng2021} have used capsule networks for the prediction of transcription factor binding sites. Their sequence encoding methods are respectively one-hot encoding and dna2vec. But the results are not directly comparable as the papers do not specifically mention the cell lines used in their work. However, the average results as reported in the respective papers are lower than that of the proposed DNABERT-Cap.

\section{Discussion}
In this work, we present DNABERT-Cap which is a transformer based capsule network to predict TFBSs. The results show that the combination of these two very powerful deep learning methods improve the prediction performance as reported in the Results section. Moreover, we have also performed ablation studies to report the utility of applying DNABERT and capsule network for such prediction. In this regard, the improved performance of the proposed model can be attributed to the ability of DNABERT embeddings to generate rich bidirectional contextual representations by having multiple attention heads focusing on various input sections concurrently. Moreover, the ability of capsule network in keeping information about the location of an object as opposed to a traditional convolutional network that loses track of it when the pooling layers only extract the most important information from data. The proposed model also benefits from the joint optimisation of DNABERT and capsule layers along with convolutional and BiLSTM layers to learn important attributes and features for TFBSs. Similar to how spatial correlation is crucial for correctly identifying objects in images, ordering of \textit{k}-mers and their semantic representations are important for DNA sequences as well. The proposed model seems to identify such relationships and thus perform better predictions.  As a future research work, attempts can be made to further improve the performance of the model by not only considering DNA sequences for feature embedding but also other parameters as well. Moreover, DNABERT-Cap can be used to address other prediction problems in bioinformatics such as predicting RNA and DNA-protein binding sites from sequences.

\section{Conclusion}
In this work we propose DNABERT-Cap for the identification of transcription factor binding sites in DNA sequences. 
%
We show here that such combination of DNABERT and capsule network may have an important role in exploiting data of high complexity. 
In order to learn characteristics specific to DNA sequences and TFBSs, DNABERT-Cap relies on the beneficial strength of combining a pre-trained DNABERT model and a capsule layer. The proposed model is also compared with  state-of-the-art approaches and shows an accuracy of more than 83\% and an AUC of more than 0.91 for all the five cell lines as considered in this work. We hope that this work will be useful for the researchers working on applications of deep learning models in bioinformatics. 

\section*{Acknowledgments}
The work presented in this paper has been supported by PNRR MUR project PE0000013-\textbf{FAIR}. 
The authors show their gratitude to the Italian CINECA Consortium for graciously providing the required computational resources.

\bibliographystyle{plainnat}
\bibliography{ref}

\begin{table*}\tiny
    \centering

    \setlength\tabcolsep{1pt}
    \begin{tabular}{|l|l|c|c|c|c|c|}\hline
        Model & Cell line & Accuracy (\%) & Recall (\%)& Specificity (\%) & MCC & AUC\\\hline
          \multirow{5}{*}{Fine-tuned DNABERT model (Baseline-1)} &A549  & 79.14$\pm$ 0.407 & 78.55$\pm$ 2.511 & 79.72$\pm$ 0.426 & 0.609$\pm$ 0.015 & 0.873$\pm$ 0.004 \\\cline{2-7}
         &GM12878  & 78.13$\pm$ 0.226 & 76.87$\pm$ 2.116 & 78.43$\pm$ 0.246& 0.608$\pm$ 0.004& 0.854$\pm$ 0.001 \\\cline{2-7}
         &Hep-G2   & 79.56$\pm$ 0.116 & 78.20$\pm$ 1.413 & 79.88$\pm$ 0.118& 0.609$\pm$ 0.004& 0.876$\pm$ 0.000 \\\cline{2-7}
         & H1-hESC & 78.11$\pm$ 0.412 & 77.14$\pm$ 3.126  & 78.13$\pm$ 0.426 & 0.603$\pm$ 0.005& 0.858$\pm$ 0.001  \\\cline{2-7}
         &Hela   & 78.19$\pm$ 0.215 & 77.98$\pm$ 1.457 & 78.56$\pm$ 0.219 & 0.604$\pm$ 0.006 & 0.873$\pm$ 0.002  \\\hline
        \multirow{5}{*}{DNABERT+CL+BiLSTM+CE (Baseline-2)} &A549  & 80.45$\pm$ 0.310 & 80.63$\pm$ 2.026  & 80.01$\pm$ 0.315 & 0.629$\pm$ 0.004& 0.890$\pm$ 0.000 \\\cline{2-7}
         &GM12878  & 80.32$\pm$ 0.167 & 78.73$\pm$ 2.551 & 80.67$\pm$ 0.171& 0.625$\pm$ 0.006 & 0.877$\pm$ 0.001 \\\cline{2-7}
         &Hep-G2  & 81.65$\pm$ 0.488 & 80.66$\pm$ 2.073 &81.37$\pm$ 0.483 &0.639$\pm$ 0.009 & 0.896$\pm$ 0.004\\\cline{2-7}
         & H1-hESC & 80.51$\pm$ 0.334 & 78.01$\pm$ 2.778  & 80.59 $\pm$ 0.339& 0.631$\pm$ 0.004&  0.892$\pm$ 0.003\\\cline{2-7}
         &Hela   & 81.64$\pm$ 0.145 & 69.02$\pm$ 1.887 & 81.55$\pm$ 0.146 & 0.643$\pm$ 0.002&  0.905$\pm$ 0.002\\\hline
           \multirow{5}{*}{DNABERT+CL+Capsule Layer(Baseline-3)} &A549  & 83.58$\pm$ 0.331 & \textbf{83.03$\pm$ 1.734} & 79.58$\pm$ 0.336  & 0.681$\pm$ 0.004& 0.915$\pm$ 0.001 \\\cline{2-7}
         &GM12878  & 82.65$\pm$ 0.177 & 77.57$\pm$ 2.374 & 82.65$\pm$ 0.181 & 0.659$\pm$ 0.008&  0.903$\pm$ 0.000\\\cline{2-7}
         &Hep-G2   & 84.76$\pm$ 0.414 & 79.92$\pm$ 1.483 & 84.68$\pm$ .417& 0.698$\pm$ 0.001& 0.920 $\pm$ 0.001\\\cline{2-7}
         & H1-hESC & 82.62$\pm$ 0.357 & \textbf{78.85$\pm$ 2.656} & 82.59$\pm$ 0.351 & 0.654  $\pm$ 0.003& 0.904$\pm$ 0.001\\\cline{2-7}
         &Hela   & 83.09$\pm$ 0.126 & \textbf{80.68$\pm$ 1.913} & 83.00$\pm$ 0.126& 0.662$\pm$ 0.002& 0.908$\pm$ 0.002 \\\hline
         \multirow{5}{*}{DNABERT-Cap} &A549  & \textbf{84.66 $\pm$ 0.302} & 81.57 $\pm$ 1.711 & \textbf{84.65$\pm$ 0.311}& \textbf{0.696$\pm$ 0.004}& \textbf{0.925$\pm$ 0.001} \\\cline{2-7}
         &GM12878  & \textbf{83.52$\pm$ 0.173} & \textbf{81.11 $\pm$ 2.110}  & \textbf{83.52$\pm$ 0.173}& \textbf{0.671$\pm$ 0.003} & \textbf{0.913$\pm$ 0.001} \\\cline{2-7}
         &Hep-G2   & \textbf{85.49$\pm$ 0.157} & \textbf{83.43$\pm$ 1.640} & \textbf{85.46$\pm$ 0.161} & \textbf{0.710$\pm$ 0.003}& \textbf{0.930$\pm$ 0.001} \\\cline{2-7}
         & H1-hESC & \textbf{83.43$\pm$ 0.350}  & 78.39$\pm$ 2.652 & \textbf{83.50$\pm$ 0.305} & \textbf{0.674$\pm$ 0.003} & \textbf{0.914$\pm$ 0.001}  \\\cline{2-7}
         &Hela   & \textbf{83.94$\pm$ 0.112} & 80.46$\pm$ 1.834 & \textbf{83.89$\pm$ 0.105}& \textbf{0.680$\pm$ 0.002} & \textbf{0.917$\pm$ 0.000}  \\\hline

    \end{tabular}
        \caption{Summary of the performances of the DNABERT-Cap and the selected baselines. The results indicate that DNABERT with Capsule Network show the best overall performance.}
    \label{tab1}
\end{table*}
\begin{table*}\scriptsize
    \centering
    
    \setlength\tabcolsep{1pt}
    \begin{tabular}{|l|c|c|c|c|}\hline
        Model & Accuracy (\%) & Recall (\%)& Specificity (\%) & MCC \\\hline
        DNABERT-Cap & \textbf{84.21} & 80.99 & \textbf{84.20 }& \textbf{0.686} \\\hline
        DeepARC        & 83.10 & \textbf{82.02} & 84.19 & 0.664\\\hline
        DeepTF      & 80.98 & 77.44 & 81.36 & 0.632\\\hline
        CNN-Zeng & 79.92 & 72.12 & 81.96 & 0.619\\\hline
        DeepBind         & 79.82 & 72.64 & 81.44 & 0.609\\\hline
        \end{tabular}
        \caption{Summary of performances of DNABERT-CAP and state-of-the-art predictors. DNABERT-Cap has the best overall performance among all the compared predictors.}
    
    \label{tab2}
\end{table*}

\end{document}